\documentclass[useAMS,usenatbib]{mn2e}

\usepackage{graphicx,epsfig}

\def\gsim{\;\lower4pt\hbox{${\buildrel\displaystyle >\over\sim}$}\;}
\def\lsim{\;\lower4pt\hbox{${\buildrel\displaystyle <\over\sim}$}\;}
\def\grls{\;\lower4pt\hbox{${\buildrel\displaystyle >\over <}$}\;}

\title[Line Profiles from Star-Forming Clouds]
{Molecular Line Profiles of Collapsing Gas Clouds}
\author[Y. Gao, Y.-Q. Lou and K. Wu]
{Yang Gao$^{1}$\thanks{
  E-mail: gaoyang-00@mails.tsinghua.edu.cn (YG);
  louyq@tsinghua.edu.cn (Y-QL);
  kw@mssl.ucl.ac.uk (KW)},
  Yu-Qing Lou$^{1,3,4}$\footnotemark[1]
  and Kinwah Wu$^{2}$\footnotemark[1] \\
$^1$Department of Physics and Tsinghua Centre for Astrophysics
  (THCA), Tsinghua University, Beijing 100084, China \\
$^2$Mullard Space Science Laboratory, University College London,
  Holmbury St. Mary, Surrey, RH5 6NT, United Kingdom \\
$^3$National Astronomical Observatories of China, Chinese Academy of Sciences,
   A20, Datun Road, Beijing 100012, China \\
$^4$Department of Astronomy and Astrophysics,
  The University of Chicago, 5640 South Ellis Avenue, Chicago, IL 60637, USA }
\date{Accepted 2009 August 5. Received 2009 July 31; in original form 2008 November 15}

\begin{document}
\maketitle

\begin{abstract}
Emission line profiles of tracer molecule H$_2$CO 140~GHz transition
  from gravitational core collapsing clouds in the dynamic
  process of forming protostars are calculated,
  using a simple ray-tracing radiative transfer model.
Three self-similar dynamic inside-out core collapse models --
  the conventional polytropic model, the empirical hybrid model and the
  isothermal model -- for star-forming molecular clouds are examined
  and compared.
The isothermal model cannot produce observed
  asymmetric double-peak molecular line profiles.
The conventional polytropic model, which gives flow velocity, mass
  density and temperature profiles self-consistently, can produce
  asymmetric double-peak line profiles for a core
collapsing cloud. In particular,
  the blue peak is stronger than the red peak,
  consistent with a broad class of molecular line profile observations.
We find that line profiles are robust against variations in the
  polytropic index $\gamma$ once the effective line-centre opacity
  $\kappa_0$
  is specified.
The relative strengths of the blue and red peaks within a molecular
  line profile are determined by the cloud temperature gradient,
  but the emission at frequencies
  between the two line peaks is determined by detailed density and velocity
  profiles in the cloud core.
In the presence of a static dense kernel at the centre of a
  collapsing cloud,
  strong internal absorption along the line-of-sight may occur,
  causing a suppression to the red wing of the blue line peak.
If reliably resolved in frequency by observations,
  this signature may be potentially useful for probing the environs
  of an infant protostar. The conventional polytropic model can be
  utilized to produce molecular line-profile templates,
  for extracting dynamical information from line spectra of
  molecular globules undergoing a gravitational core collapse.
We show a sample fit using the 140~GHz H$_2$CO emission line from
  the central region of the molecular globule B335 by our model with $\gamma=1.2$.
The calculation of line profiles and fitting processes also offer a
  scenario to estimate the protostellar mass, the kernel mass accretion rate,
  and the evolution time scale of a core collapsing cloud. Our model can be
  readily adapted to other tracer molecules with more or less constant abundances
  in star-forming clouds.
\end{abstract}


\section{Introduction}

How molecular clouds collapse to form protostars
  is an important question in modern astronomy and cosmology.
Since hydrodynamic collapse models for star formations in molecular
  clouds were first proposed in the 1960s
  \citep[e.g.][]{BS1968,larson1969,penston1969},
  substantial progress has been achieved in modelling dynamic
  properties of core collapsing clouds
  \citep[e.g.][]{shu1977,hunter1977,tsai1995,lou2004,ShenLou04,
  fatuzzo2004,ShenLou06,lou2006,myers2008,hulou08,evans2009,louzhai09}.
These progressive research advancement leads us to a more
  realistic understanding of
  the star formation processes in molecular clouds.
In particular, the gas flow velocity, mass density and thermal
  temperature structures of collapsing clouds predicted by theoretical
  models are in principle testable by various high-resolution spectral
  imaging observations in different bands of tracer molecular line
  profiles \citep[e.g.][]{snell1981,zhou1993,mardones1997,yang2002,
  devries2002,saito1999,belloche2002,evans2005}.

A widely discussed scenario for the formation of low-mass protostars
  is the so-called `inside-out collapse' model
  \citep[][]{shu1977}.
In this model, a condensed cloud core becomes unstable
  gravitationally and collapses rapidly towards the centre to form a
  protostellar core.
The protostar is embedded in a dusty envelope, which is presumed
  to be static, although in some variations of the model the
  envelope may also involve an infall or an outflow
  \citep[e.g.][]{lou2004,ShenLou04,hulou08,louzhai09}.
The boundary between the collapsing core and the envelope expands,
  engulfing more and more mass into the central collapsing core.
The accretion of dusty materials from the envelope forms a disc
  rotating about the central protostar, and magnetized bipolar
  outflows may occur from the disc at the same time \citep[e.g.][]{shu1987}.
The inside-out model has enjoyed supports from observations of some
  molecular globules.
The Bok molecule globule B335
  \citep[e.g.][]{frerking1987,zhou1993,choi1995,saito1999,stutz2008}
  serves as a test example among others \citep[e.g.][]{hogerheijde2000}.
However, there are features of line emissions from this
  star forming cloud \citep[see e.g.][]{wilner2000} and also
  observations of other cloud systems, such as IRAM~04191
  \citep[e.g.][]{andre1999,belloche2002} and L1544
  \citep[e.g.][]{tafalla1998,vandertak2005}, that are not readily
  explained by the {\it isothermal} inside-out collapse scenario as
  advanced by Shu et al. (1987)

Different from the inside-out collapse scenario for a molecular
  cloud, there are alternative models to describe the formation
  of low-mass protostars within molecular clouds, e.g. the isothermal
  Bonnor-Ebert sphere embedded in an external medium of finite pressure
  $P_{\rm E}$ \citep{bonnor1956,ebert1955}. This model describes the
  hydrostatic equilibrium of a gaseous sphere with a constant kinetic
  temperature, while a central core with finite density is involved.
  Such a hydrostatic sphere can be maintained as long as the mass of
  the cloud does not exceed the critical value of
  $M_{\rm c}=1.18a^4/(G^{3/2}P_{\rm E}^{1/2})$, where $a$ is the isothermal
  sound speed and $G$ is the gravitational constant. With density profile
  and kinetic temperature being the two key aspects, dynamic instability
  is predicted therein, but no flow velocity is presumed. Recent research
  shows that the hydrostatic Bonnor-Ebert sphere may
  match several observed properties of starless cores
  \citep[e.g.][]{evans2001,belloche2002,myers2005},
  while it cannot account for the observational data of B335 source
  \citep[see e.g.][]{harvey2003}.
This isothermal sphere model with a finite mass in a hydrostatic
  equilibrium cannot fully describe observations of molecular line
  profiles in the cloud core because of the very isothermal assumption
  as we shall show presently in the following sections.

The Larson-Penston (LP) self-similar solution
  \citep{larson1969,penston1969} of
  an isothermal sphere is a classic dynamic model for low-mass
  star formation.
This LP type model predicts the formation of stellar density cores
  around the centre of molecular clouds. Characterisitcally,
  the decreasing-towards-centre radial collapse speed profiles
  is obtained along with corresponding mass density profiles.
However, the isothermal assumption encounters the similar
  problem of the Bonnor-Ebert sphere for modelling the observed
  molecular line profiles.
Relaxing the dynamic self-consistency, all these models may be
   partially applied semi-empirically to star formation clouds for
   developing diagnostics. For example,
recent empirical works make use of the millimetre continuum
   emission and provide empirical thermodynamic (density and
   temperature) profiles
   \citep[e.g.][]{tafalla2002,shirley2002,evans2005}. These
model results offer an independent perspective to probe
   star-forming clouds, but the overall dynamic profiles lack the
   desired self-consistency. In reference to these dynamic models,
   current observations and empirical works, the polytropic EECC
   solution \citep[e.g.][Lou \& Hu 2009]{louwang06,lou2006,wang2008}
   appears to be a sensible alternative.
With radial velocity fields, corresponding density profiles and
  variable temperature profiles, this self-similar dynamic model
  should be more practical and versatile in explaining different
  observations (Lou \& Gao 2006).

Molecular emission lines provide valuable observational diagnostics
  to probe the dynamics and physical conditions in star-forming clouds
  \citep[e.g.][]{DysonWilliams1997}.
For examples, optically thick emission lines of tracer molecules
  H$_2$CO, HCO$^+$, CO, CS and N$_2$H$^{+}$ are frequently identified
  for such diagnostic purposes.
However, there are still unknown and uncertain parameters as well
  as model degeneracies to explain the line strengths and profiles, due
  to global hydrodynamics, macro- and micro-scale kinematics, thermal
  structures, chemical compositions and turbulence.
It is thus important to resolve these issues step by
  step to avoid ambiguities in spectral modelling as well as in both
  observational and theoretical interpretations.

To date, most molecular line profile calculations assume a certain
  parametrization for the density, velocity and temperature structures
  within emitting clouds.
For example, in \citet{zhou1993}, the so-called `exact theory' takes
  the density and velocity profiles from the {\it isothermal} dynamic
  model of \citet{shu1977} but the temperature profile is assigned
  separately with a certain empirical base and therefore the gas cloud
  model is actually {\it non-isothermal} for molecular line
  calculations.
This is a worthwhile attempt, yet such a specification of the radial
  temperature structure is unsatisfactory as there are inconsistencies
  in the velocity and density profiles of an {\it isothermal} cloud
  from the theoretical perspective.
In recent years, calculations \citep[e.g.][]{devries2005}
  considered more structural complexities in star-forming molecular
  clouds and invoked a more sophisticated radiative transfer model
  that incorporates the relevant atomic processes \citep[e.g.][]{ward2001,
  rawlings2001,keto2004,tafalla2006,tsamis2008,pavlyuchenkov2008}.
These studies aimed at resolving specific issues, such as the
  chemical profiles \citep[e.g.][]{rawlings2001,tsamis2008} or the
  competition between relevant atomic processes
  \citep[e.g.][]{pavlyuchenkov2008}.
As the effects of dynamical and thermal structures were not their
  prime objective, most of these calculations basically parameterized
  the density, velocity and temperature structures in a collapsing
  cloud or simply adopted the isothermal inside-out collapse model of
  Shu (1977) but took an empirical temperature profile from
  observations of dust continuum.

The key question now is: does it make a difference
  if the structure profile of a cloud is determined self-consistently
  through a polytropic hydrodynamic model analysis instead of using a simple
  isothermal inside-out model but with a non-isothermal
  temperature profile?
Also, are the molecular line profiles different
  for models with different polytropic indices of the gas dynamics?
These questions are essential for our understanding of the dynamics
  and physical conditions, in particular, in the early stage of
  forming low-mass protostars.
The objective of our study here is to address this issue
  quantitatively. For instance,
we calculate H$_2$CO 140~GHz molecular line profiles using a
  self-similar hydrodynamic model that involves a conventional
  polytropic equation of state (EoS) \citep[Fatuzzo et al. 2004;]
  [Hu \& Lou 2008]{
  louwang06,lou2006,wang2008}
  and an isothermal inside-out collapse model \citep{shu1977},
  and carefully compare their differences.
We also directly compare the H$_2$CO 140~GHz molecular line profiles
  obtained from the conventional polytropic model
  \citep{lou2006} with observations, using the molecular cloud B335
  as a test example.
Comparison of our theoretical approach with empirical approaches
  used in former works is also made. In principle, our model
  calculations can be readily adapted to line profiles of tracer
  molecules of more or less constant abundances in star-forming clouds.

This paper is structured as follows.
In Section 2, we present model structures of conventional polytropic
  and isothermal inside-out collapse models with their flow velocity,
  density and temperature profiles.
The relevant parameters and scalings of dynamic models are also
  specified there. We provide an overview for the formation of
  molecular line profiles in star-forming clouds in this section.
In Section 3, we present the results of calculations,
  the comparison of the lines obtained by dynamical models
  and the comparison of the theoretical with the observed
  molecular line profiles.
Effects due to the cloud temperature gradient, optical depth,
  intrinsic line width, polytropic index and the presence of a
  relatively dense central kernel are also considered.
Finally, results of our analysis are summarized in Section 4. For
  the convenience of reference, we summarize the basic
  polytropic hydrodynamics and the relevant asymptotic self-similar
  solutions in Appendix A, elaborate on how the blue and red peaks
  of a molecular line profile are shaped in an optically thick cloud
  in Appendix B, and lay out the radiative transfer formulation
  and our ray-tracing computation algorithm in Appendix C.

\section{Self-Similar Hydrodynamic Models}

%
\begin{figure}
\begin{center}
\epsfig{figure=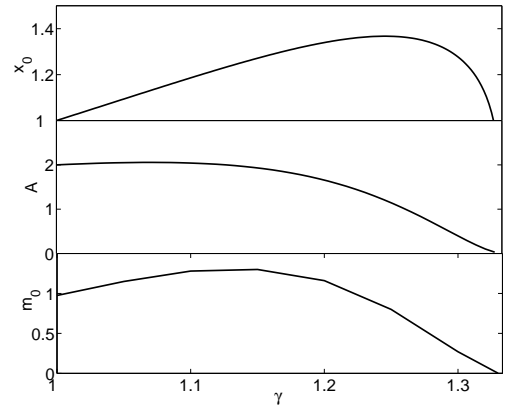,width=8.5cm,clip=}
\end{center}
\caption{Parameter variations as functions of polytropic index
  $\gamma=2-n$ for the expansion-wave collapse scenario with static
  SPS outer envelopes and dynamic free-fall cores as in Model I.
The upper panel is the infall radius $x_0$,
  the middle panel is the mass parameter $A$
  given by eq. (\ref{equ:A}) for a SPS solution (\ref{Equ:static})
  and the bottom panel is the point mass $m_0$ for $x\rightarrow 0^{+}$.
Consistently, $\gamma=1$ is the isothermal case (Model II)
  with parameters $x_0=1$, $A=2$ and $m_0=0.975$.
All these parameters increase slightly and then fall as $\gamma$
  increases from $1$ to $(4/3)^{-}$.
The special case of $\gamma=4/3$ ($n=2/3$) is investigated by Lou
  \& Cao (2008) and Cao \& Lou (2009).
  \label{fig:dynamicvariable}}
\end{figure}

Three self-similar hydrodynamic models for the velocity, density
  and temperature profiles are considered in our molecular
  line-profile calculations, namely
  Model I, a hydrodynamic collapse model with a conventional polytropic
  EoS for the gas \citep[Hu \& Lou 2008]{lou2006};
  Model II, an isothermal inside-out collapse model \citep{shu1977};
  and Model III, an empirical hybrid model \citep{zhou1993}
  \footnote{The hybrid model of Zhou et al. (1993) takes the mass
  density and flow velocity profiles from the isothermal model of
  Shu (1977) and uses an empirical variable temperature profile.
  This empirical temperature profile was inferred based on an
  isothermal density profile which scales as $r^{-2}$.}.
In all these three models, the molecular cloud consists of a
  free-fall collapsing inner core and a static outer envelope.
A self-similar dynamic
  solution is adopted for the collapsing core in both Model I and II.
For Model I,
  we further use three specialized solutions for comparison.
In fact, the isothermal case (Model II) can be regarded as a
  very special case of the conventional polytropic model (Model I)
  with $\gamma=n=1$.

\subsection{Parameters of Hydrodynamic Model Solutions}

Details of
  Model I and relevant asymptotic solutions are contained in
  Appendix A for reference.
Model I is characterized by $n+\gamma=2$ where
  $n$ is a scaling index for polytropic self-similar hydrodynamics
  and $\gamma$ is the polytropic index (see Appendix A; Lou \& Wang
  2006; Lou \& Gao 2006; Hu \& Lou 2008).
We choose different dynamic solutions with a static outer envelope
  and free-fall inner core collapse for comparison in terms of
  tracer molecule line profiles.
The density, velocity
  and temperature profiles
  can be directly derived from equations (\ref{Equ:function1}),
  (\ref{Equ:function2}), (\ref{equ:temp}) for a conventional polytropic
  gas dynamics under self-gravity.

\begin{table}
 \centering
 \begin{minipage}{100mm}
  \caption{Parameters of self-similar polytropic dynamic solutions
  \label{Table:dynamic}}
  \begin{tabular}{@{}llllll@{}}
  \hline
   $\gamma$ & $n$ & $x_0$ & $m_0$ & $A$ & $B$ \\
 \hline
 1.0 & 1.0 & 1.00 & 0.975& 2.00 & 0 \\
 1.1 & 0.9 & 1.17 & 1.25 & 2.04 & 0 \\
 1.2 & 0.8 & 1.35 & 1.15 & 1.66 & 0 \\
 1.3 & 0.7 & 1.28 & 0.26 & 0.40 & 0 \\
\hline
\end{tabular}
\end{minipage}
\end{table}

For conventional polytropic solutions with $\gamma=1.1$,
  $1.2$ and $1.3$, the corresponding values of the dimensionless
  infall radius $x_0$, the dimensionless core mass $m_0$, and
  coefficients $A$ and $B$ for asymptotic solution(\ref{Equ:infinity})
  are contained in Table \ref{Table:dynamic}.
Parameters for the isothermal model
  of $\gamma=n=1$ are also listed in Table \ref{Table:dynamic}.
In our constructed solutions with static singular polytropic
  solution (SPS) envelopes, the dimensionless infall radius $x_0$
  depends on $n$ only and is analytically given by
  \begin{equation}
  x_0=\frac{(2-n)^{1/2}}{n}[2(3n-2)]^{1-n/2}\ .\label{equ:x_0}
  \end{equation}
The velocity parameter $B$ of asymptotic solution for static
  envelope vanishes and the mass parameter $A$ can be derived
\begin{equation}
A=\bigg[\frac{2(3n-2)(2-n)}{n^2}\bigg]^{1/n}\ ,\label{equ:A}
\end{equation}
  which is exactly the same as the coefficient of $\alpha(x)$ in SPS
  solution (\ref{Equ:static}).
The dimensionless central point mass $m_0$ of asymptotic free-fall
  solution (\ref{Equ:zero1})
  is then determined numerically.
Fig. \ref{fig:dynamicvariable} shows parameter variations
  of $x_0$, $A$ and $m_0$ as the scaling index $\gamma$ (or
  equivalently $n=2-\gamma$) varies continuously.
This is in fact the case (i.e. expansion-wave collapse
  solution: EWCS) of a static SPS envelope connected with
  a free-fall collapsing core and is much more general than
  the special isothermal case of \citet{shu1977}.


\begin{figure}
\begin{center}
\epsfig{figure=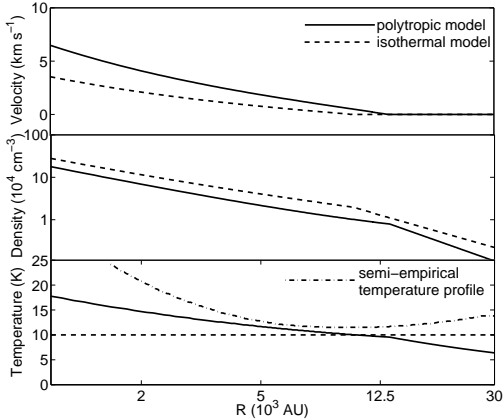,width=8.5cm,clip=}
\end{center}
\caption{Inward radial flow velocity,
  number density
  and thermal temperature
  profiles (panels from top to bottom
  respectively) of three different model molecular clouds used in our
  molecular line-profile calculations.
The abscissa represents the radius $R$ in unit of $10^3$~AU, shown
  on a logarithmic scale.
The solid curves denote Model I with $\gamma=1.2$;
  the dashed curves denote Model II.
The dash-dotted curve in the bottom panel denote Model III,
  whose redial velocity and number density has the same manner
  as Model II (dashed curves in the top and middle panels).
Other relevant model parameters can be found in Table 1 of
  $\gamma=1.0$ for Model II and III,
  and $\gamma=1.2$ for Model I.
The dimensional scalings used are described in Section 2, giving
  infall radii of $R=10\times 10^3$ AU for Model II and III,
  and $R=13.5\times 10^3$ AU for Model I ($\gamma=1.2$), respectively.
\label{fig:dynamicdata}}
\end{figure}

The dimensionless variables are converted to corresponding
  dimensional physical variables in order to perform radiative
  transfer calculations for molecular line profiles in core
  collapsing clouds.
We adopt parameter values estimated from observations for the
  scalings of the dynamic profiles for model clouds.
The typical infall radius within a molecular cloud is
  estimated as $\sim 0.01 - 0.03$~pc \citep[e.g.][]{myers2005},
  corresponding to $\sim 10^3-10^4$~AU.
As dimensionless infall radii are roughly around
  and larger than $x_0=1$ in our model (e.g. $1\lsim x_0\lsim 1.4$ by
  Fig. \ref{fig:dynamicvariable} and Table 1), we may choose the
  following length scale
  \begin{equation}
  k^{1/2}t^n=10^4~{\rm AU}\ \label{scalingOne}
  \end{equation}
  in definition (\ref{equ:varx}).
A typical number density at the infall radius is $\sim 10^4~{\rm
  cm}^{-3}$ \citep[e.g.][]{harvey2003,evans2009}.
For the mass parameter $A\lsim 2$ as shown in Fig.
  \ref{fig:dynamicvariable}, the reduced number density being of order
  of unity implies a proper number density scale as (see equation
  \ref{equ:varu})
  \begin{equation}
  1/(4\pi G\mu m_{\rm H}t^2)\cong 10^4~{\rm cm}^{-3}\ .\label{time}
  \end{equation}
This corresponds to a thermal temperature of
  $T\sim 4$~K by equation (\ref{equ:temp}) and a characteristic
  infall age of $t\sim 3\times 10^5$~years by equation (\ref{time})
  for a molecular cloud \citep[see e.g.][]{myers2005}.
Within the infall radius, this scaling also gives an
  enclosed mass scale of $M\simeq 0.5~M_{\odot}$
  and a sound speed scale
  $c\simeq 0.15~{\rm km~s}^{-1}$ by equations (\ref{equ:varu}) and
  (\ref{equ:soundspeed}).

In addition to the central protostar, we presume a number density
  of $\sim 10^6$ cm$^{-3}$ at the very central region (i.e. $R \lsim
  500\ $AU) of a molecular cloud and set the central velocity to be
  zero for a static dense kernel.
The outer cloud radius is set to be several tens of $10^3$ AU. We
  note that the size of the static dense kernel and the outer radius
  of the cloud are not the same as used in some previous works
  \citep[e.g.][]{harvey2003,myers2005,evans2005}.
The key parameter in our radiative transfer calculations
  is the effective optical depth
  of a molecular line in a cloud.
Once the numerical value of the optical depth along the ray is
  specified, the actual size of an emitting cloud would be irrelevant.
In this sense, the difference between cloud sizes used in our
  model analysis and those of earlier research works should not affect
  our basic conclusions. Nevertheless, we should keep in mind that this
  conclusion is made under the assumption that the abundance of tracer
  molecule remains constant in our model consideration. The cloud sizes
  would matter if the fractional abundances of certain
  tracer molecules vary considerably in star forming clouds.

\begin{figure}
\begin{center}
\epsfig{figure=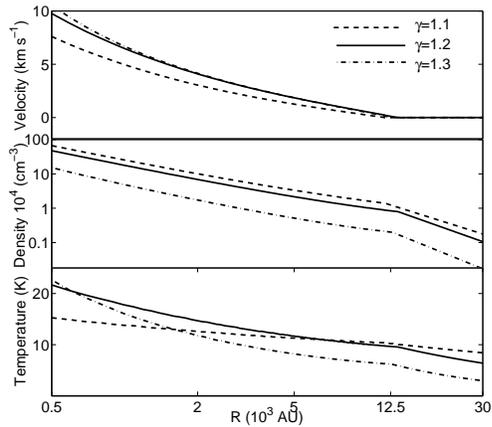,width=8.5cm,clip=}
\end{center}
\caption{Inward radial flow velocity,
  number density
  and thermal temperature
  profiles (panels from top to bottom respectively) of Model I
  with three different polytropic indices $\gamma$.
The dashed, solid, and dash-dotted curves represent $\gamma=1.1$,
  1.2, and 1.3 with infall radii at $R=11.7\times 10^3$, $13.5\times
  10^3$, and $12.8\times 10^3$ AU, respectively.
The corresponding central mass points are 1.13, 1.02 and 0.24
  $M_{\odot}$.
Other relevant polytropic model parameters are contained in Table 1
  and the physical scalings are described in Section 2.
The case of
  $\gamma =1.2$ is the same as shown in Fig.~\ref{fig:dynamicdata}.
A larger $\gamma$ provides steeper density and temperature increases
  towards the centre; this effect will manifest in the corresponding
  H$_2$CO 140GHz line profiles as discussed in Section 3.3.
\label{fig:polydata}}
\end{figure}

For comparison, the velocity, number density and temperature
  profiles of the three different models that we use in radiative
  transfer calculations of molecular line profiles are shown in
  Fig.~\ref{fig:dynamicdata}.
In all models, the velocity and temperature increase monotonically
  towards the centre until $R\approx 500$ AU.
There are slight differences between the density and velocity
  profiles of Model I ($\gamma=1.2$) and Model II.
The infall radii for Model II and Model I
  ($\gamma=1.2$) are at $R=10$ and 13.5 AU respectively.
While the temperature profiles of the three models
  are readily distinguishable.
Model II has a flat constant temperature,
  Model I ($\gamma=1.2$) has a monotonic increase of temperature towards the centre,
  and Model III has a semi-empirical profile
  in which the temperature drops slightly from
  $R=30\times 10^3$~AU to $R\approx 6\times 10^3$~AU
  and then increases when $R$ decreases further.
We also show in Fig.~\ref{fig:polydata} the velocity, density and
  temperature profiles for Model I with different polytropic indices:
  $\gamma = 1.1$, 1.2 and 1.3,
  whose infall radii are at $R= 11.7$, 13.5 and 12.8 $\times 10^3$ AU
  respectively.
As shown, cloud structures can vary substantially even within Model
  I (in fact, Model II can be regarded as a special case of Model I),
  and such differences can leave signatures in the emission molecular
  line profiles as expected.

\subsection{A Molecular Line Profile Calculation}

We utilize a simple ray-tracing radiative transfer model (which
  does not involve an explicit treatment of atom and molecular
  transitions and their couplings with the radiation field; see
  Appendix C for details) to calculate the spectral profile of
  molecule H$_2$CO (2$_{12}$-1$_{11}$) emission line at frequency
  140~GHz. As the range of temperature variation in a cloud region
  is typically small (e.g. $\sim 10-20$ K), the variation in tracer
  molecular level populations is small and thus will affect line
  profiles little \citep[e.g.][]{tsamis2008}.

It has been argued that gradients in distributions of chemical
  elements as well as chemical evolution are important in
  determining the strengths and profiles of molecular emission
  lines from star-forming clouds
  \citep[e.g.][]{rawlings2001,tsamis2008,pavlyuchenkov2008}.
As we focus on the effect of dynamical cloud structures on tracer
  molecule line profiles, the H$_2$CO abundance is taken to be
  approximately constant in our model calculations. This
  simplification implies that our approach may be extended to those
  tracer molecules with more or less constant abundances in star
  forming clouds. This assumption of constant abundance was invoked
  in some earlier works \citep[e.g.][]{zhou1993,ward2001},
  while in others, the H$_2$CO abundance in B335 cloud was assumed
  to be a step function
  with a variation of $\sim 10$ times \citep[e.g.][]{evans2005}.
In fact, recent empirical works based on observations do show that
  several diagnostic tracer molecules (e.g. HO, H$_2$CO, CN, CS)
  keep their abundance variations less than a factor of $\sim 10$
  \citep[e.g. Fig. 3 in][]{tsamis2008}.
Both theoretically and observationally, clarifications of the
  relative importance of dynamical and chemical effects are
  essential in using molecular line profiles as a reliable
  diagnostic probe to the core collapse dynamics in star-forming
  clouds.
A more comprehensive description of star-forming clouds should
  involve self-consistent dynamic profiles and chemical abundance
  profiles that are independently inferred from observations.

In order to have an intuitive feeling for our model approach,
  we demonstrate in Appendix B how the blue and red peaks of emission
  molecular line profiles are formed in an optically thick cloud.
The radiative transfer analysis and the computing algorithm of our
  line profile calculations are described in Appendix C.
The physical variables (i.e. temperature $T$, mass density $\rho$,
  projected velocity component $v_{\rm p}$ etc.) of a core
  collapsing molecular cloud are determined from our polytropic
  hydrodynamic model (see Appendix A).
The line-centre opacity $\kappa_0$ and the line width $\Delta v$ are
  set to values typical of spectral line observations of tracer molecules.
In equations (\ref{RTeq})$-$(\ref{equ:taucell}), the mass density
  $\rho$ is mainly that of the interstellar medium in a cloud.
As a consequence, the ratio of molecular abundance is merged into
  the parameter $\kappa_0$, which is then effectively a product of
  molecular line centre opacity and its mass fractional ratio.

A simple assumption of local thermodynamic equilibrium (LTE) is made
  for the radiative transfer calculation; this calculation does not
  account for scattering dominated processes, masers, or other non-LTE
  cases.

\section{Results and Analyses}

\subsection{Temperature Variations in Molecular Clouds}

\begin{figure}
\begin{center}
\epsfig{figure=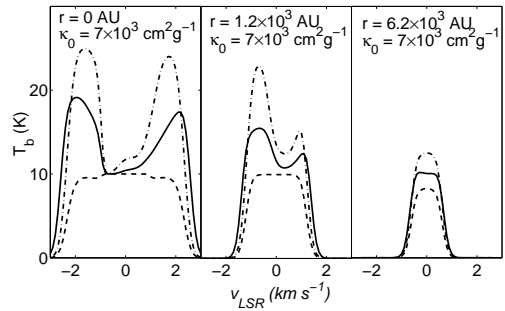,width=8.5cm,clip=}
\end{center}
\caption{
Profiles of H$_2$CO 140~GHz emission lines in terms of brightness
  temperature T$_{\rm b}$
  from molecular clouds for Model I with $\gamma=1.2$ (solid curves),
  Model II (dashed curves) and Model III (dash-dotted curves).
The radial flow velocity, particle number density and thermal
  temperature profiles of the underlying dynamic models are shown in Fig.
  \ref{fig:dynamicdata} with a central kernel to avoid the
  singularity.
The intrinsic line widths $\Delta v$ are set to
  be $\sim 0.3~{\rm km~s}^{-1}$ here.
Panels from left to right correspond to line emission profiles along
  the light-of-sight passing through
  $R=0$~AU, $R=1.2\times 10^3$~AU and $R=6.2\times 10^3$~AU, respectively.
The effective total line-centre opacity is $\kappa_0=7\times
  10^3\hbox{ cm}^2\hbox{ g}^{-1}$ in all three panels.
  \label{fig:spectra4}}
\end{figure}

It is clear that Model II does not produce asymmetric double-peak
  H$_2$CO 140~GHz line profiles (see the dashed curves in
  Fig.~\ref{fig:spectra4} and Equations~(\ref{diff_intensity})
  and (\ref{diff_temperature}) in Appendix B), an apparent
  contradiction to extensive line profile observations in
  molecular globules.
Therefore, unless compromising the self-consistency in the thermal
  structure and dynamics to insert a semi-empirical temperature profile
  to replace the isothermal model, as in Model III,
  one cannot have a sensible comparison between the observed asymmetric
  double-peak line profiles of star-forming clouds and the line profiles
  calculated from Model II.
In fact, equation (\ref{diff_temperature}) in Appendix B
  clearly demands that the temperature should
  increase towards the cloud centre in order to produce a molecular line profile
  with a blue peak being stronger than the red peak for a core collapsing cloud.
Fig.~\ref{fig:spectra4} \citep[see also Fig.~9a of][]{zhou1993}
  demonstrates that asymmetric molecular line profiles with a stronger blue peak
  emerge when Model III and Model I ($\gamma=1.2$) are actually introduced.
Recent radiative transfer works use empirical temperature profiles
  derived from the sub-millimeter continuum emission intensity are
  able to produce the observed asymmetric line profiles
  \citep[e.g.][]{ward2001,evans2001,tsamis2008}.
However, self-consistency of the dynamic profiles should be
  demanded for a more satisfying model construction.

  We emphasize that the
  temperature profile is determined simultaneously once we derive mass
  density and flow velocity profiles from conventional polytropic
  hydrodynamic equations. Instead, when one uses an empirical temperature
  profile for molecular line profile calculations, this empirical
  temperature profile and the density and velocity profiles are generally
  not consistent with each other in the hydrodynamic and thermodynamic sense.
  In other words, a gas cloud with dynamic profiles having no self-consistency
  does not exist according to hydrodynamics and thermodynamics. It should be
  also noted that the inference of empirical temperature profiles must also
  rely on several assumptions. Our goal is to develop theoretical models
  grossly satisfying available observational constraints.

Our calculations show that in both Model I ($\gamma=1.2$) and Model
  III, the asymmetry in line profiles can be more pronounced in the
  off-centre region in a cloud, provided the ray of emission along
  the line-of-sight passes through the collapsing core of the cloud
  (compare the left and middle panels of Fig.~\ref{fig:spectra4}).
  The relative strengths of the blue and red peaks in a molecular
  line profile show a stronger contrast in Model III than
  in Model I ($\gamma=1.2$).
This sharper contrast in two peaks from Model III is due to the
  fact that there are more variations in the specified semi-empirical
  temperature profile than those in the temperature profile of
  Model I ($\gamma=1.2$) that we adopt.
Model II produces symmetric single-peak molecular line profiles in
  the resolved spectra by our numerical explorations (see three
  dashed curves of Fig.~\ref{fig:spectra4}), while Model I ($\gamma=1.2$)
  and Model III could produce single peak line profiles when the rays are
  along the line-of-sight passing through only the static outer envelope
  or when velocity variations are insignificant in the emission region
  as compared with the intrinsic velocity broadening $\Delta v$ of
  H$_2$CO 140~GHz line.

\subsection{Line Opacity and Intrinsic Velocity Width}

\begin{figure}
\begin{center}
\epsfig{figure=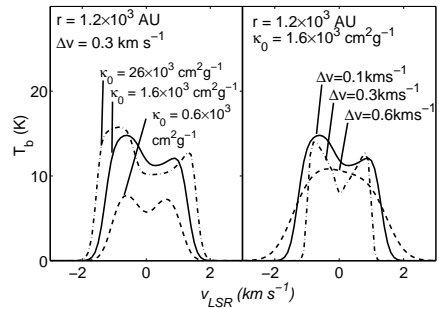,width=8.0cm,clip=}
\end{center}
\caption{
Profiles of H$_2$CO 140~GHz emission lines in terms of brightness
  temperature T$_{\rm b}$
  from molecular clouds at $R=1.2\times 10^3$~AU for Model I
  of $\gamma=1.2$
  with different values of the effective line-centre opacity
  $\kappa_0=0.6\times 10^3\ ,\ 1.6\times 10^3\ ,\ 2.6\times 10^4
  \hbox{ cm}^2\hbox{ g}^{-1}$ (left panel), and for different values of
  intrinsic line widths $\Delta v$ (0.1, 0.3 and 0.6~${\rm km~s}^{-1}$)
  with a fixed $\kappa_0=1.6\times 10^3\hbox{ cm}^2\hbox{ g}^{-1}$
  (right panel).
Unless otherwise stated, the parameters adopted are
   the same as those in Model I in Fig. \ref{fig:dynamicdata}.
\label{fig:spectrad}}
\end{figure}

The left panel of Fig.~\ref{fig:spectrad} shows how the two peaks
  in H$_2$CO 140~GHz line profile vary with the value of the line
  centre opacity $\kappa_0$.
With density distribution being fixed, the opacity $\kappa_0$
  characterizes the optical depth
  along the ray parallel to the line-of-sight.
For forming asymmetric blue and red peaks,
  molecular line profiles need to be optically thick,
  in addition to the presence of variable temperature and velocity
  structures within molecular clouds involving gravitational core collapse.
For molecular lines that are optically thin,
  the observed emissions sample the entire collapsing core.
A cloud structure of spherical symmetry for the near and far
  hemispheres inevitably gives rise to symmetric molecular line
  profiles, in spite of the temperature, density and velocity
  variations in the emission region.
Moreover, as the line is optically thin,
  the line intensity is well below the source Planck function.
Hence the line intensity becomes weaker,
  i.e. the corresponding brightness temperature is lower than the
  thermal temperature of line emission regions
 (see equation \ref{diff_temperature} and discussions thereafter
  for an intuitive explanation for the formation of a double-peak
  line profile).

For comparisons of dynamic models, we have adopted that the
  intrinsic line width $\Delta v =0.3~{\rm km~s}^{-1}$.
There are uncertainties in the values of $\Delta v$ and we have
  chosen a typical value here.
It is generally believed that the intrinsic width of a molecular
  line is mainly caused by thermal broadening and micro-turbulence.
The thermal broadening is estimated to be at a level of $\sim
  0.1~{\rm km~s}^{-1}$ for a typical molecular cloud with a thermal
  temperature of $\sim 10$~K.
  The turbulent broadening is fairly uncertain.
Theoretical estimates depend on the assumed driving mechanisms
  \citep[e.g.][]{maclow1999,li2006,matzner2007}.
Some observational studies \citep[e.g.][]{larson1981} show that the
  probable line width $\Delta v$ ranges from a few $\sim 0.1~{\rm
  km~s}^{-1}$ to about several ${\rm km~s}^{-1}$.

Typically, infall velocities $v$ in cloud cores would range
  from $\sim 0.1~{\rm km~s}^{-1}$ to a few ${\rm km~s}^{-1}$,
  as adopted in the dynamic models explored here.
The characteristic double-peak signature of molecular line profiles
  from a collapsing cloud may be erased or smeared if the line suffers
  a severe turbulent broadening (i.e. large $\Delta v$).
Right panel of Fig~\ref{fig:spectrad} demonstrates that for Model I
  ($\gamma=1.2$) with a maximum infall velocity of $\sim 2~{\rm km~s}^{-1}$,
  we can produce a double-peak line profile for
  $\Delta v\lsim 0.3~{\rm km~s}^{-1}$ but not for $\Delta v
  \gsim 0.6~{\rm km~s}^{-1}$.
Extensive observations of double-peak line profiles in many
  molecular globules imply that the turbulence in their line
  formation region should not give rise to velocity broadenings
  significantly larger than $\sim 0.3~{\rm km~s}^{-1}$
  with a maximum infall velocity $\sim 2~{\rm km~s}^{-1}$.
By a further numerical exploration,
  we find that the maximum allowed velocity broadening bears
  a linear relationship with the maximum infall speed;
  the values shown here are typical for star-forming clouds.


\subsection{Polytropic Index $\gamma$}

One appealing aspect of Model I is that we can obtain
  self-consistent velocity, density and temperature profiles for
  molecular clouds within the overall inside-out collapse scenario.
The essence of this model is the conventional polytropic equation of
  state characterized by a polytropic index $\gamma$.
The special isothermal case (Model II) just corresponds to
  $\gamma=n=1$.
Now our question is: how robust is the molecular line profile
  morphology with respect to the variation of $\gamma$?

\begin{figure}
\begin{center}
\epsfig{figure=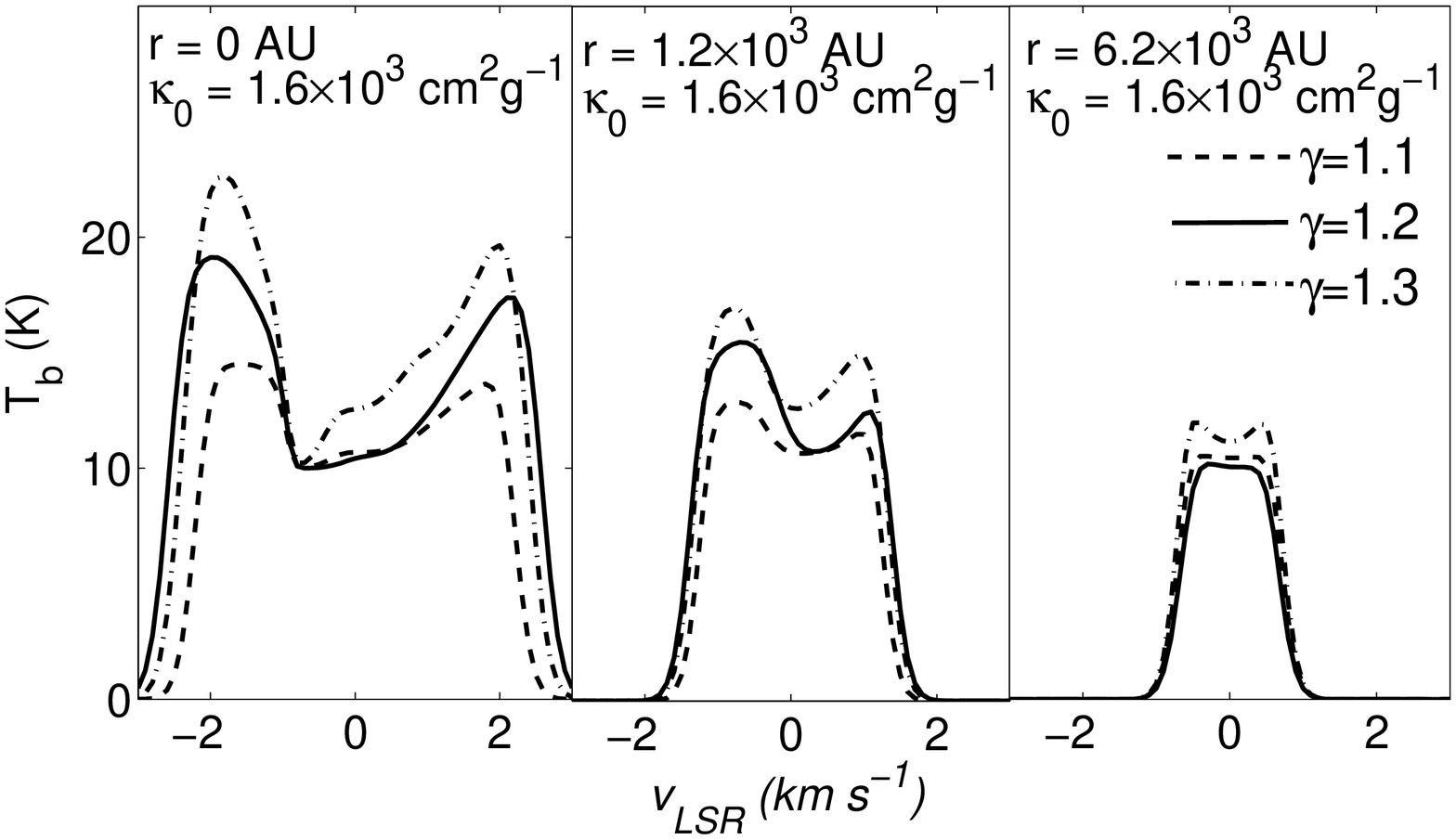,width=8.5cm,clip=}
\end{center}
\caption{Profiles of H$_2$CO 140~GHz emission lines in terms of
  brightness temperature T$_{\rm b}$
  from molecular clouds for the three polytropic models shown in
  Fig.~\ref{fig:polydata}.
The dash-dotted curve corresponds to the model with $\gamma=1.1$;
  the solid curve to the model with $\gamma=1.2$; and the dashed
  curve to the model with $\gamma=1.3$.
The scale free infall radii are at $x_0=1.18,\ 1.35$ and 1.27
  respectively.
For all three panels, $\kappa_0=7\times 10^3\hbox{ cm}^2\hbox{
  g}^{-1}$.
\label{fig:spectrapoly}}
\end{figure}

We calculate H$_2$CO 140~GHz emission line profiles using Model I
  with
  three different polytropic indices $\gamma=1.1$, 1.2 and 1.3.
These $\gamma$ values are representative of astrophysical
  situations between limiting regimes, e.g. the isothermal process of
  $\gamma=1$, and the adiabatic processes of $\gamma=7/5$ and $5/3$
  for diatomic and monoatomic gases respectively.
We found that the overall morphology of molecular line profiles is
  quite robust with respect to variations of $\gamma$,
  provided that the effective line-centre opacities in the calculations
  take on the same values (see Fig. ~\ref{fig:spectrapoly}).
Note that $n+\gamma=2$ for a conventional polytropic gas, i.e. a
  variation of $\gamma$ leads to a variation of $n$,
  which is related to the scalings of hydrodynamic variables.
There are however some small and subtle differences among predicted
  molecular line profiles.
For instance, larger $\gamma$ tends to give more prominent peaks in
  a line profile,
  which can be attributed to a much steeper gradient in the temperature
  increase towards the cloud centre (Fig. ~\ref{fig:polydata}).
Also, for a larger $\gamma$ value, the molecular line peaks persist
  to larger projected radii away from the
  cloud centre (see the right panel of Fig. ~\ref{fig:spectrapoly}).
Nevertheless, these differences might not be readily distinguishable
  in observations, given the beam smearing in the spectral imaging
  observations and the presence of various noises in molecular line
  spectra.

\subsection{The Effect of a High-Density Central Kernel}

In theoretical models of gravitational core collapsing clouds,
  a static central dense kernel is introduced to avoid
  the singularity.
Phenomenologically, the radius of the kernel marks the interface of
  an `embryo' stellar kernel with the accreting gas in a gravitational
  free fall.
This high-density kernel is more a theoretical hypothesis as
  was used in former works \citep[e.g.][]{choi1995,myers2005}
  but it does fit the observed sub-millimeter intensity profile well
  \citep[e.g.][]{evans2001}.

\begin{figure}
\begin{center}
\epsfig{figure=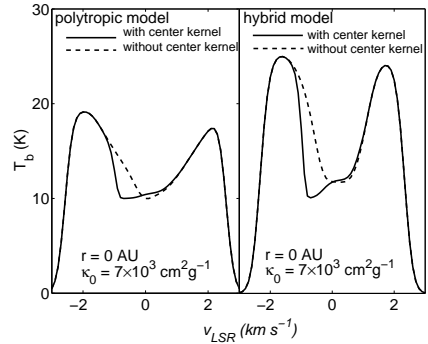,width=8.0cm,clip=}
\end{center}
\caption{Comparison of H$_2$CO 140~GHz emission line profiles in
  terms of the brightness temperature T$_{\rm b}$ along the
  light-of-sight passing through the cloud centre for
  Model I ($\gamma=1.2$) (left panel) and Model III (right panel).
The difference between the two model curves within each
  panel is that the solid curve is the same as the cloud
  as in Fig.~\ref{fig:dynamicdata} and dashed curve has the
  innermost zero-velocity condensed kernel removed.
Other parameters of the `kernel-less' model clouds are the same
   as those in Model I and Model III in Fig. \ref{fig:dynamicdata}.
In both panels, $\kappa_0=7\times 10^3\hbox{ cm}^2\hbox{ g}^{-1}$.
\label{fig:centerspectra}}
\end{figure}

The static high-density kernel could leave indelible signatures in
  molecular line profiles from the central core of a collapsing cloud.
As the effective kernel opacity is large,
  molecular emissions from the far side (i.e. behind the dense kernel)
  would suffer strong absorptions at the rest-frame line-centre
  frequencies, depending on intrinsic line broadening $\Delta v$.
In contrast, molecular emissions from the near side (i.e. in front
  of the dense kernel) is unaffected by the kernel.
This naturally leads to the suppression of the red wing of the blue
  line peak.
In the left panel of Fig. \ref{fig:spectra4}, molecular
  line profiles manifest a dip feature in the red wing of the blue
  peak when the line-of-sight passes through the dense kernel.
For rays along the lines-of-sight that avoid this central
  dense kernel, this dip feature would disappear (see the middle panel
  of Fig. \ref{fig:spectra4} as an example).
Fig.~\ref{fig:centerspectra} compares molecular line profiles
  from both Models I ($\gamma=1.2$) and III with and without a central kernel.
The difference in molecular line profiles for the case with and
  without the dense static kernel are not negligible.
Moreover, the presence of a dip makes the red peak appearing
  broader than the blue peak.
However, this difference in line profiles is similar to the
  difference caused by different values of $\gamma$ in Model I
  (the left panel of  Fig.\ref{fig:spectrapoly}),
  as well as the difference between Model I and Model III
  (left panel of  Fig.\ref{fig:spectra4}).
So the actual cause of this phenomenon on profiles can only be
  solved after testing the temperature and density profiles of
  the cloud by some other means (e.g. the millimetre continuum).

Potentially, such a line feature may be utilized to probe the
  `embryo' kernel surrounding a protostar in its development and
  evolution [e.g. pre-protostellar cores (PPCs) with detectable dust
  continuum emission at millimetre and sub-millimetre wavelengths]
  before the eventual emergence as a `fully functional' star of
  nuclear burning.
However in reality, whether or not such dips may be unambiguously
  resolved in the observed line profiles from molecular globules
  is determined by a few factors. In particular, it depends on
  the spectral resolution actually achieved by observations.

\subsection{Comparisons with Observed Line Profiles}

It is straightforward that self-similar solutions of Model I
  combined with radiative transfer calculations (Appendix C)
  can be applied to fit observed H$_2$CO 140~GHz line profiles.
Empirical estimates of physical variables at certain epochs
  are introduced for calibration and profiles of variables
  are derived from Model I self-consistently.
As an example of illustration, we specifically model the 140~GHz
  H$_2$CO emission line from the Bok molecular globule B335.
We take the observed line profile from \citet{zhou1993},
  where the line emission was extracted along the line-of-sight
  path towards the cloud centre.
The collapsing core (i.e. the infalling region around
  the centre) radius is $\sim 30$" according to \citet{zhou1993}.
By adopting an estimated distance of $\sim 250$~pc to the Bok
  globule B335 \citep[][]{tomita1979}, this gives a radius of
  $\sim 7.5\times 10^3$~AU for the infalling region.
However, the uncertainty in this distance to B335 is fairly large
  (see Table 2 of Tomita et al. 1979)
  and this radius may vary by more than a factor of two.
We take this radius to be $\sim 1\times 10^4$ AU for the
  collapsing core in our model calculations for B335 cloud.

\begin{figure}
\begin{center}
\epsfig{figure=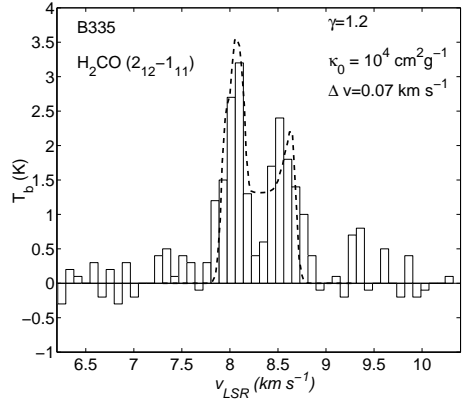,width=8.50cm,clip=}
\end{center}
\caption{Comparison of the 140~GHz H$_{2}$CO ($2_{12}-1_{11}$)
  line profile towards the central region of the Bok molecular
  globule B335 with the theoretical molecular line profile obtained
  by Model I with $\gamma=1.2$ (dashed line).
This abundance ratio of H$_{2}$CO molecule is taken to be $7\times
  10^{-9}$ and $\kappa_0=10^4\hbox{ cm}^2\hbox{ g}^{-1}$.
The radial receding displacement velocity of the entire cloud is
  $v_{\rm d}\sim 8.30~{\rm km~s}^{-1}$.
The infall speed is $\sim 0.27~{\rm km~s}^{-1}$ at a radial distance
  of $\sim 10^3$~AU away from the cloud centre.
The gas temperature of the central kernel is $\sim 3.6$~K.
\label{fig:spectra2}}
\end{figure}

The same $\gamma=1.2$ solution of Model I as in Fig.
  \ref{fig:dynamicdata} is used to model the structure
  of molecular line formation region.
The maximum infall velocity is taken to be $\sim 0.50~{\rm
  km~s}^{-1}$
  at $0.6\times 10^3$~AU and the radial receding displacement velocity
  $v_{\rm d}$ of the entire B335 cloud is $\sim 8.30~{\rm km~s}^{-1}$.
  The intrinsic width of the molecular line is $\sim 0.07~{\rm
  km~s}^{-1}$, and the effective line-centre opacity is
  $\kappa_0=10^4\hbox{ cm}^2\hbox{ g}^{-1}$.
We find that a dense static inner kernel is needed in order to
  produce line profiles similar to the observed profiles.
The kernel radius is $\sim 0.5\times 10^3$~AU and the mean
  particle number density in the kernel is $\sim 10^6~{\rm cm}^{-3}$.
Fig.~\ref{fig:spectra2} compares the observed line profile
  and the line profile produced by Model I ($\gamma=1.2$).
The good agreement between our model line profile and the observed
  line profile clearly shows the viability of the conventional
  polytropic hydrodynamic model and the applicability of our
  line-profile radiative transfer calculations.

It is possible to derive the central point mass $M_0$ and the
  total enclosed mass contained within the central kernel region
  $M_k$ in Model I used in the above data fitting.
Using the $m_0$ value (see Table \ref{Table:dynamic} and Fig.
  \ref{fig:dynamicvariable}) and scaling relations (\ref{scalingOne})
  and (\ref{time}) in our dynamic model with self-similar
  transformation (\ref{equ:varu}),
  we obtain the value of central point mass $M_0=1.025 M_{\odot}$,
  referred to as an estimate for the mass of the central protostar;
  this estimate is about twice the value of \citet{zhou1993}
  and \citet{stutz2008} and is $\sim 10$ times larger than the
  mass value estimated by \citet{saito1999}.
When we use the reduced enclosed mass $m(x)$ at the boundary of the
  central kernel, the dimensional enclosed mass contained within the
  kernel region is derived to be $M_k=1.027 M_{\odot}$.
Estimates of all these masses are not sensitive to the kernel
  radius but are mainly affected by the variation in the infall radius.
Therefore, all these masses would have an uncertainty of a few
  times. It is also possible to derive the mass accretion rate in the
  periphery around the kernel,
  $\dot{M}_k=0.8\times 10^{-6} M_{\odot}$ yr$^{-1}$.
In this case, the timescale of a protostar formation can then be
  estimated as $t=M_k/\dot{M}_k\sim 1.25\times 10^6$ yr, which is
  $\sim 10$ times the infall timescale estimated in Model III
  \citep{zhou1993}.


\section{Conclusions}

We have investigated and analyzed effects of radial dynamical
  and thermal structures of an inside-out core collapsing cloud
  with spherical symmetry on forming molecular line profiles
  emitted from star-forming molecular clouds.
A simple radiative transfer model employing a direct ray-tracing
  algorithm along the line of sight is constructed for computing
  H$_2$CO 140~GHz emission line profiles in order to compare with
  those actually observed in a star-forming cloud.
We model the gravitational core collapse of a molecular cloud by the
  self-similar hydrodynamics of a conventional polytropic gas under
  self-gravity in a self-consistent manner (Model I).
We emphasize that an {\it isothermal} inside-out collapse model
  (Model II) cannot produce asymmetric molecular line profiles.
In contrast, a variable radial temperature structure,
  which is crucial for determining asymmetric molecular line profiles,
  can be derived self-consistently together with the radial
  velocity and density structures in Model I,
  which is different from the semi-empirical approach (Model III).
We explicitly demonstrate that Model I can indeed produce asymmetric
  double-peaked line profiles with stronger blue peaks than red peaks
  as observed in many star-forming molecular globules.
We assess the relative importance of effects due to line-centre
  opacities, intrinsic line widths
  and polytropic index $\gamma$ in Model I.
Our radiative transfer calculations further show that the presence
  of a dense static inner kernel around the centre of a core collapsing
  cloud gives a characteristic signature of a suppressed red wing of the
  blue line peak in the molecular line profile.
We show an application to line profile observations by modelling the
  140~GHz H$_2$CO emission line profile from the centre of the Bok
  molecular globule B335.
We note that sensible fits to the H$_2$CO emission line
  profile of B335 requires a dense static kernel in addition to
  conditions of a polytropic core collapse model and an estimated
  line-of-sight receding speed of the entire molecular cloud.
Using Model I and the molecular line profile fits,
  we could estimate the protostellar mass,
  the mass inside the static condensed kernel,
  the mass accretion rate just outside the kernel radius,
  and the time scale of the protostar formation.

\section*{Acknowledgments}

This research was supported in part by Tsinghua Centre for
  Astrophysics (THCA), by the National Natural Science Foundation
  of China (NSFC) grants 10373009 and 10533020 at Tsinghua University,
  and by the Yangtze Endowment and the SRFDP 20050003088 and
  200800030071 at Tsinghua University.
The hospitality of Institut f\"ur Theoretische Physik und
  Astrophysik der Christian-Albrechts-Universit\"at Kiel Germany and of
  International Center for Relativistic Astrophysics Network (ICRANet)
  Pescara, Italy is gratefully acknowledged. KW thanks G. Sarty for comments.

\appendix

\section{\quad Hydrodynamics of a Conventional Polytropic Gas}

For the hydrodynamics of a conventional polytropic gas model of
  spherical symmetry without diffusive effects, we use the basic
  nonlinear fluid equations in spherical polar coordinates
  ($r,\ \theta,\ \phi$), namely
  \begin{equation}
  {{\partial\rho}\over{\partial t}}
  +{1\over{r^2}}{{\partial}\over{\partial r}}(r^2\rho u)=0\ ,
  \label{Equ:mass1}
  \end{equation}
  \begin{equation}
  {{\partial u}\over{\partial t}}+u{{\partial u}\over {\partial
  r}}=-{1\over{\rho}}{{\partial p}\over {\partial
  r}}-{{GM}\over{r^2}}\ , \label{Equ:force}
  \end{equation}
  \begin{equation}
  {{\partial M}\over{\partial t}} +u{{\partial M} \over{\partial
  r}}=0\ , \qquad\qquad\quad {{\partial M}\over{\partial r}} =4\pi
  r^{2}\rho\ .\label{Equ:mass2}
  \end{equation}
Here the mass density $\rho$, radial bulk flow velocity $u$,
  thermal pressure $p$, and enclosed mass $M$ are all functions of
  radius $r$ and time $t$,
  and $G=6.67\times 10^{-8} \hbox{ g}^{-1}
  \hbox{ cm}^{3}\hbox{ s}^{-2}$ is the gravitational constant.
Equations (\ref{Equ:mass1}) and (\ref{Equ:force}) are the mass
  and radial momentum conservations; equation (\ref{Equ:mass2})
  represents another form of mass conservation; the Poisson
  equation relating $\rho$ and the gravitational potential
  $\Phi$ is consistently satisfied by the above equations.

The polytropic equation of state reads
  \begin{equation}
  p=K(t)\rho^\gamma\ ,\label{Equ:state}
  \end{equation}
  where $\gamma$ is the polytropic index and $K(t)$ is a
  coefficient.
The self-similar dynamics involves an independent variable
  \begin{equation}
  x=r/(k^{1/2}t^n)\ , \label{equ:varx}
  \end{equation}
  where $n$ is a scaling index and $k$ is referred
  to as the sound parameter, and the corresponding
  self-similar transformation of other relevant
  hydrodynamic variables are
  \begin{eqnarray}
  \rho=\frac{\alpha(x)}{(4\pi Gt^2)}\ ,\qquad\quad
  M=\frac{k^{3/2}t^{3n-2}m(x)}{(3n-2)G}\ ,
  \qquad\qquad\quad\nonumber\\
  u=k^{1/2}t^{n-1}v(x)\ ,\qquad\
  p=kt^{2n-4}\alpha^{\gamma}(x)/(4\pi G)\ ,\quad\label{equ:varu}
  \end{eqnarray}
  where $\alpha(x)$, $m(x)$ and $v(x)$ are dimensionless reduced
  mass density, enclosed mass and radial flow velocity, respectively.
In reference to equation of state (\ref{Equ:state}),
  we identify $K(t)\equiv k(4\pi G)^{\gamma-1}t^{2(\gamma+n-2)}$.
For $n+\gamma=2$, $K$ becomes a constant for a conventional
  polytropic
  gas \citep[e.g.][]{sutosilk1988,louwang06,lou2006,hulou08}.
The gas temperature $T$ and the local polytropic sound speed $c$ are
  \begin{equation}
  T\equiv \frac{\mu m_{\rm H}}{k_{\rm B}}\frac{p}{\rho}= \frac{\mu m_{\rm
  H}}{k_{\rm B}} \alpha^{\gamma-1} k t^{2n-2}\ ,\label{equ:temp}
  \end{equation}
  \begin{equation}
  c\equiv (\partial p/\partial\rho)^{1/2}
  =(k\gamma)^{1/2}t^{n-1}\alpha^{(\gamma-1)/2}\ ,\label{equ:soundspeed}
  \end{equation}
  where $k_{\rm B}$, $\mu$ and $m_{\rm H}$ represent Boltzmann's
  constant, mean molecular weight and the hydrogen mass, respectively.
For typical star-forming clouds, we shall take $\mu\cong 1$ for
  simplicity.
This self-similar transformation deals with an important
  subset of nonlinear solutions to polytropic hydrodynamic partial
  differential equations (PDEs) (\ref{Equ:mass1})$-$(\ref{Equ:state})
  and provides the dynamic structures for radiative transfer calculations
  of molecular line profiles.

With self-similar transformation (\ref{equ:varu}) in nonlinear
  PDEs (\ref{Equ:mass1})$-$(\ref{Equ:state}),
  we obtain two coupled nonlinear ordinary
  differential equations (ODEs)
  \begin{eqnarray}
  {\big[(v-nx)^2-\gamma\alpha^{\gamma-1}\big]\frac{dv}{dx}}
  =\frac{(v-nx)^2\alpha}{(3n-2)}\qquad\qquad\nonumber\\
  \qquad\qquad+\frac{2(v-x)\gamma\alpha^{\gamma-1}}{x}-(v-nx)(n-1)v\
  ,
  \label{Equ:function1}
  \end{eqnarray}
  \begin{eqnarray}
  \big[(v-nx)^2-\gamma\alpha^{\gamma-1}\big]\frac{d\alpha}{\alpha
  dx}=
  (n-1)v \qquad\qquad\nonumber\\
  -\frac{(v-nx)\alpha }{(3n-2)}-\frac{2(v-x)(v-nx)}{x}\ .\label{Equ:function2}
  \end{eqnarray}
By setting $v=0$ for all $x>0$, we obtain a static singular
  polytropic solution (SPS) from equations (\ref{Equ:function1})
  and (\ref{Equ:function2})
  \begin{eqnarray}
  v=0\ ,\qquad\qquad
  \alpha=\big[{2\gamma(3n-2)}/{n^2}\big]^{{1}/{n}}
  x^{-{2}/{n}},\nonumber\\
  \vbox{\vskip 0.5cm}
  m=(2\gamma)^{{1}/{n}}(3n-2)^{{1}/{n}}
  n^{-{\gamma}/{n}}x^{{(4-3\gamma)}/{n}}\
  \ .\label{Equ:static}
  \end{eqnarray}
To leading orders,
  asymptotic similarity solutions of coupled nonlinear ODEs
  (\ref{Equ:function1}) and (\ref{Equ:function2}) for
  $x\rightarrow +\infty$ and $x\rightarrow 0^{+}$ are
  summarized below.
In the limit of $x\rightarrow +\infty$, we have
  \begin{eqnarray}
  \alpha=Ax^{-{2}/{n}}\ ,\qquad v=-\frac{nA}{(3n-2)}x^{{(n-2)}/{n}}
  \qquad\qquad \nonumber\\
  \quad\qquad +\frac{2\gamma A^{\gamma-1}}{n}
  x^{{(2-2\gamma-n)}/{n}}+Bx^{{(n-1)}/{n}}\ ,\label{Equ:infinity}
  \end{eqnarray}
  where $A$ and $B$ are two constants of integration.
In asymptotic solution (\ref{Equ:infinity}), it is possible
  to set $B=0$ with a finite $A>0$ such that $v$ vanishes at large
  $x$; and this $A$ is precisely the coefficient of $\alpha(x)$ in
  SPS solution (\ref{Equ:static}).
This consistency is expected but SPS solution (\ref{Equ:static}) is
  valid for all $x$ except for the central singularity.
In the limit of $x\rightarrow 0^{+}$,
  the asymptotic central free-fall solution is
  \begin{equation}
  v=-\bigg[\frac{2m_{0}}{(3n-2)x}\bigg]^{1/2}\ ,\qquad
  \alpha=\bigg[\frac{(3n-2)m_{0}}{2x^{3}}\bigg]^{1/2}
  \label{Equ:zero1}
  \end{equation}
  where $m=m_0$ is the reduced enclosed mass $m(x)=\alpha x^2(nx-v)$
  as $x\rightarrow 0^{+}$, for a point mass at the centre.

By free-fall solution (\ref{Equ:zero1}) and transformation
  (\ref{equ:varu}) with $n+\gamma=2$, we have a variable central
  mass accretion rate $\dot{M}$
  given by
  \begin{equation}
  \dot{M}=k^{3/2}t^{(3-3\gamma )}m_0/G\ .\label{massaccretion}
  \end{equation}
Only for the isothermal case of $\gamma=1$ ($n=1$),
  can we have a constant $\dot{M}$.
We solve nonlinear ODEs (\ref{Equ:function1}) and
  (\ref{Equ:function2}) numerically using the fourth-order Runge-Kutta
  method.
The SPS solution (\ref{Equ:static}) and asymptotic conditions
  (\ref{Equ:infinity}) and (\ref{Equ:zero1}) are all used to determine
  self-similar dynamic flows.
We also pay attention to the sonic critical curve (see Lou \& Gao
  2006, Lou \& Wang 2006 and Hu \& Lou 2008 for details)

\section{\quad Formation of Double-Peak Molecular Line Profiles}

Extensive spectral observations often show two peaks in high
  excitation emission molecular line profiles from star-forming
  molecular globules, and the blue peak is usually stronger than the
  red peak (i.e. blue asymmetric lines).
This phenomenon can be explained in the framework of the inside-out
  collapse model \citep{shu1977}.
We recount below the relevant molecular line profile formation
  processes \citep{zhou1993}.

\begin{figure}
\begin{center}
\epsfig{figure=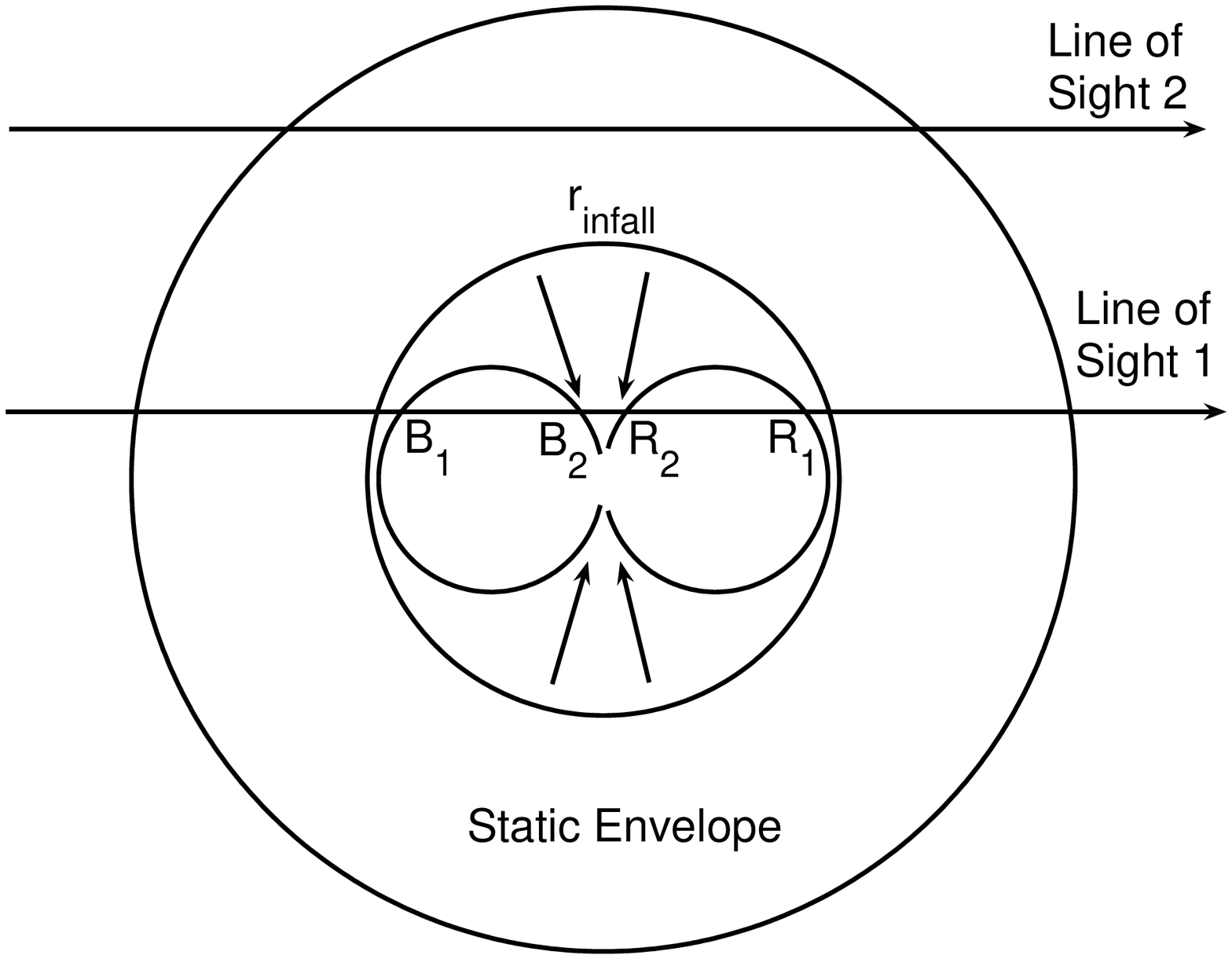,width=8.0cm,clip=}
\end{center}
\caption{A schematic illustration of the inside-out collapse
  scenario for the formation of double-peaked molecular line profiles.
The cloud consists of an outer static envelope and an inner
  collapsing core, separated by a spherical surface at $r=r_{\rm
  infall}$ in radial expansion.
The loci (i.e. the two connected ovals) of a particularly chosen
  constant projected line-of-sight velocity component are shown.
Two lines of sight are drawn towards right, labelled by `Line of
  Sight 1' and `Line of Sight 2', respectively.
Line of Sight 1 passes through the core of central infall as well as
  the outer static envelope,
  that intercepts the ovals at points B$_1$ and B$_2$ for
  the same approaching projected velocity components and at points
  R$_1$ and R$_2$ for the same receding projected velocity components.
Sufficiently far away from the centre, Line of Sight 2 passes
  through only the outer static envelope.\label{fig:schematic1}}
\end{figure}

Consider a spherically symmetric and non-rotating molecular cloud.
In the isothermal inside-out collapse model of \citet{shu1977},
  the cloud consists of two radial regions, viz. (i) an inner
  collapsing core and (ii) an outer static envelope, as depicted in
  Fig.~\ref{fig:schematic1}.
For simplicity, we use a static envelope here.
The dynamical infall at the inner region determines the velocity
  profiles and hence the families of loci with equal magnitudes of
  projected line-of-sight velocity components.
In Fig.~\ref{fig:schematic1}, we show two loci of the families
  that have equal magnitudes of projected line-of-sight velocity
  components but opposite directions.
In the limit of no absorption (i.e. the optically thin regime),
  molecular lines from the emitters on these two loci will split,
  forming two symmetric components of peaks with equal but opposite
  (blue and red) frequency shifts.

Let us first pick up Line of Sight 2 which passes
  through only the outer static envelope.
This is a simple situation that the line has no bulk velocity
  broadening and splitting, i.e. a single-peaked line profile
  characterized by thermal and microscopic turbulence broadenings
  would be expected.
Next we consider Line of Sight 1 which passes through both the outer
  static envelope and the infalling core.
Line of Sight 1 intercepts the two loci of each family of constant
  projected velocity component at four points,
  namely B$_1$, B$_2$, R$_1$ and R$_2$,
  with B and R implicating blue and red shifts respectively.
We may denote the intensities of emission line components at
  these blue and red shifts as $I_{\rm blue}$ and $I_{\rm red}$.
Suppose that the absorption of the emission line by the
  continuum is insignificant in the collapsing core.
For illustration here, we  may ignore the extinction in the
  envelope without losing generality, as it introduces only a
  scaling factor equally applied to both $I_{\rm blue}$ and
  $I_{\rm red}$.
With this simplification, the emission intensities are given by two
  transfer equations,
  \begin{equation}
  I_{\rm blue} = S_1(1-e^{-\tau_1})e^{-\tau_2}+S_2(1-e^{-\tau_2}), \
  \label{blue}
  \end{equation}
  \begin{equation}
  I_{\rm red} =  S_2(1-e^{-\tau_2})e^{-\tau_1}+S_1(1-e^{-\tau_1}),
  \label{red}
  \end{equation}
  \citep[e.g.][]{zhou1993},
  where in the vicinity of points 1 and 2, $\tau_{1}$ and $\tau_{2}$ are
  the optical depths, while $S_{1}$ and $S_{2}$ are the source functions,
  with subscripts 1 and 2 denoting the outer and inner portions within the
  collapsing core at which four points (B$_1$, R$_1$) and (B$_2$, R$_2$)
  are located.
It follows that the difference between the blue and red components
  of the line at this projected velocity component is
  \begin{equation}
  I_{\rm blue}-I_{\rm red}=(S_2-S_1)(1-e^{-\tau_2})(1-e^{-\tau_1})\
  . \label{diff_intensity}
  \end{equation}
For a local thermal equilibrium (LTE), the source function $S$ is
  given by the local Planck function \citep[e.g.][]{Rybicki1979}.
In the Rayleigh-Jeans regime of lower frequencies, the spectral
  intensity $I_{\nu}$ may be expressed in terms of the brightness
  temperature T$_{\rm b}$ in the form of $I_{\nu}=2\nu^2k_B{\rm
  T}_{\rm b}/c^2$, where $\nu$, $c$ and $k_B$ are the frequency, the
  speed of light and Boltzmann's constant, respectively
  \citep[e.g.][]{DysonWilliams1997}.
Hence the difference in brightness temperatures between the blue and
  red peak emissions of a molecular line profile is simply
  \begin{equation}
  T_{\rm b,blue} - T_{\rm b,red} = (T_2 -T_1)(1-e^{-\tau_2})(1-
  e^{-\tau_1})\ .\label{diff_temperature}
  \end{equation}
Equation~(\ref{diff_temperature}) shows that a maximum contrast in
  the two peaks happens in the regime of
  $\tau_1\gg 1$ and $\tau_2\gg 1$, and this gives a brightness
  temperature peak difference $T_{\rm blue}-T_{\rm red}\approx T_2-T_1$.
Asymmetric line profiles arise only when there are substantial
  line optical depths in the hemispheres of the collapsing core
  that give rise to blue and red shifts of the line.
In the opposite regime of $\tau_1\ll 1$ and $\tau_2\ll 1$,
  the brightness temperature peak difference becomes
  $T_{\rm blue}- T_{\rm red}\approx (T_2 -T_1)\tau_1\tau_2\approx 0$.
In this case, the equal strengths of the two peaks can be understood
  as a direct consequence that in the absence of absorption (i.e. an
  optically thin gas) the emission from any location along the
  line-of-sight is visible.

Molecular emission lines commonly used for cloud structure
  diagnostics
  \citep[e.g.\ molecules H$_2$CO, HCO$^{+}$, CS, N$_2$H$^+$ etc.
  see][]{zhou1990,zhou1993,DysonWilliams1997,ward2001,lee2004,tsamis2008}
  are opaque in molecular globules that form protostars.
For a collapsing core under the self-gravity, an increase in the
  temperature towards the centre, i.e. $T_2>T_1$,
  would then lead to a stronger blue peak than the
  red peak in the emission line profile.
The observed stronger blue peaks in the H$_2$CO and CS line profiles
  from the molecular globule B335
  thus support the notion of a collapsing core and in particular, the
  inside-out collapse scenario.
  But such clouds cannot be isothermal.
  Moreover, the temperature should be higher towards the cloud
  centre.

\section{\quad Radiative Transfer Calculations of Molecular Lines}


The ray-tracing radiative transfer equation in the LTE
  approximation reads
  \begin{equation}
  \frac{d}{ds}I_\nu =-(\kappa_{\rm c}+\kappa_{\rm l})\rho I_\nu
  + \kappa_{\rm c}\rho S_{\rm c}
  + \kappa_{\rm l}\rho S_{\rm l}\ ,\label{RTeq}
  \end{equation}
  \citep[e.g.][]{Chandra1960} where $ds$ is the ray path differential element,
  $\rho$ is the mass density, $S_{\rm c}$ and $S_{\rm l}$ are the source
  functions, and $\kappa_{\rm c}$ and $\kappa_{\rm l}$ are the mass absorption
  coefficients (or opacities),
  and the subscripts `c' and `l' refer to associations with the continuum and
  line emissions respectively.
The corresponding optical depths are defined by
  \begin{equation}
  \tau_{\rm c}=\int_0^s ds'\rho\kappa_{\rm c}(s')
  \qquad\hbox{ and }\qquad
  \tau_{\rm l}=\int_0^s ds'\rho\kappa_{\rm l}(s')\ .
  \label{equ:tau}
  \end{equation}
For spectral line emissions, the opacity takes the form of
  $\kappa_{\rm l}=\kappa_{\rm l_0}\phi_\nu$, where $\kappa_{\rm l_0}$
  is the line centre opacity and dimensionless $\phi_\nu$ is the
  normalized spectral line-profile function in radiation frequency
  $\nu$.
The continuum remains more or less constant in the narrow frequency
  range of a spectral line profile;
  thus a molecular line profile is mainly determined by the opacity,
  source function and the intrinsic line profile function $\phi_{\nu}$.
For simplicity, we ignore continuum contributions by setting
  $\kappa_{\rm c} = 0$ and $S_{\rm c}=0$ at this stage.
Hereafter and in the main text, subscript `l' is also suppressed.

\begin{figure}
\begin{center}
\epsfig{figure=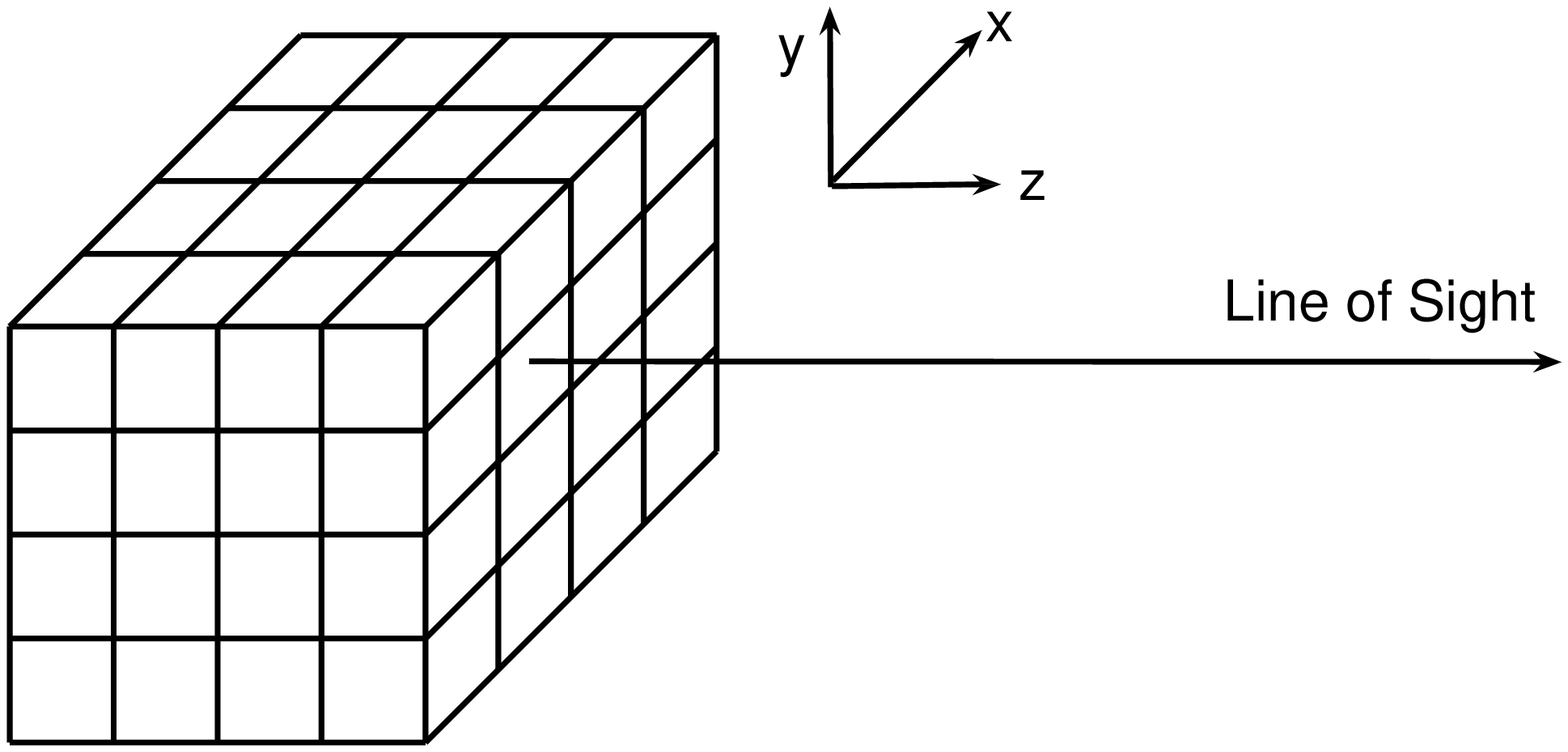,width=8.25cm,clip=}
\end{center}
\caption{An illustration of the computational
  grid in our ray-tracing radiative transfer
  calculations (with $4^3=64$ cubic cells).
The intensity of a molecular line is calculated cell by cell
  sequentially along each ray parallel to the $z$ axis.
  The spectral image is formed in the $(x,\ y)$ plane of sky.
In our calculations for molecular emission line profiles,
  typically $101^3$ cubic cells are used.\label{fig:gridfigure}}
\end{figure}


To solve radiative transfer equation (\ref{RTeq}),
  we apply a forward ray-tracing computational algorithm
  on a three-dimensional (3D) Cartesian $(x,y,z)$ grid with
  $N^3$ cells with $N$ being the number of grids along each side.
  A spherical molecular cloud is embedded inside such a cubic grid.
A ray propagates parallel to the $z-$direction towards an observer,
  and the spectral image of a cloud is projected onto the $(x,\ y)$ plane
  of sky perpendicular to the $z-$direction (see Fig.~\ref{fig:gridfigure}).
We use $N^3=101^3$ cubic cells with an equal edge length of
  $\Delta s$ for each grid cell which provide a sufficient spatial
  resolution for the purpose of our line profile computations.

Approximately, each cell is locally homogeneous in density,
  temperature and velocity, with their values taking the
  local values of a model molecular cloud at the cell centre.
The numerical difference scheme for the local solution to transfer
  equation (\ref{RTeq}) in a cell can be cast as
  \begin{equation}
  I_{\rm n}=I_{\rm n-1} e^{-\Delta\tau_{\rm n}}+S_{\rm n}
  (1-e^{-\Delta\tau_{\rm n}})\ ,\label{equ:transfer}
  \end{equation}
  where $I_{\rm n}$ is the emerging emission from the n-th cell, and
  $S_{\rm n}$ and $\Delta\tau_{\rm n}$ are the source function and the
  optical depth of the n-th cell.
By equation (\ref{equ:tau}), the optical depth of the n-th cell is
  \begin{equation}
  \Delta \tau_{\rm n}=\rho\kappa\Delta s =\rho\kappa_0 \phi(v,\
  v_{\rm p})\Delta s\ ,\label{equ:taucell}
  \end{equation}
  where $\rho$ is the mass density of the cell, $\kappa$ is the
  effective line opacity, $\kappa_0$ is the line-centre opacity and
  $\phi(v,\ v_{\rm p})$ is the intrinsic normalized line profile.
$\phi(v,\ v_{\rm p})$ in equation (\ref{equ:taucell}) is a
  function of velocity components $v$ and $v_{\rm p}$, which
  correspond to the emission frequency $\nu$ and the gas velocity in
  the cell projected along the line-of-sight respectively.
We assume for simplicity that molecular line profiles are
  intrinsically Gaussian with a normalized line profile function in
  the form of
  \begin{equation}
  \phi_\nu\rightarrow\phi(v,\ v_{\rm p})=
  (2\pi\Delta v^2)^{-1/2}\exp
  \left[ -\frac{(v-v_{\rm p})^2}{2\Delta v^2}\right]\ ,
  \label{equ:kappa}
  \end{equation}
  where $v_{\rm p}$ is the velocity component of the emitter
  projected along the line-of-sight and $\Delta v^2$ is the
  square of the `intrinsic' characteristic broadening of a
  tracer molecular line in the absence of large-scale
  systematic dynamic flows.
This intrinsic broadening includes both the thermal and
  microturbulence effects.
The source function of the n-the cell is given by the local
  Planck function $(2h\nu^3/c^2)\{\exp[h\nu/(k_BT)]-1\}^{-1}$ where
  $h$ is the Planck constant.
For emission lines in the Rayleigh-Jeans regime of low frequencies,
  we then have $S_{\rm n}=2\nu^2k_BT/c^2$,
  where $T$ is the thermal temperature of the cell.

Our radiative transfer calculations are first carried out in a
  cell for a fixed frequency $\nu$ and proceed sequentially to the
  adjacent cell along a ray parallel to the $z-$direction.
Such calculations are then repeated for a range of frequencies along
  the same ray path.
After finishing the integration along a ray, the same procedure will
  be repeated for all other rays on the grid,
  which then generates a molecular line spectrum for each spatial
  pixel on the image plane.
Depending on the desired spatial resolution, summation over the
  spectra of relevant pixels would give a line spectrum from a part of
  a molecular cloud.

\end{document}